\documentclass[conference]{IEEEtran}
\IEEEoverridecommandlockouts
\usepackage{cite}
\usepackage{amsmath,amssymb,amsfonts}
\usepackage{algorithmic}
\usepackage{graphicx}
\usepackage{textcomp}
\usepackage{multirow}
\usepackage{booktabs}
\usepackage{xcolor}

\def\BibTeX{{\rm B\kern-.05em{\sc i\kern-.025em b}\kern-.08em
    T\kern-.1667em\lower.7ex\hbox{E}\kern-.125emX}}

\begin{document}

\title{Graph-Spectral Fusion of Wavelet Packets and Higher-Order Statistics for Anomaly Detection in Industrial IoT Networks\\
{\footnotesize
\thanks{This work is supported by the US-Ireland R\&D Partnership Programme Project ``Resilient Networks'' under Grant RI-SFI-23/US/3924, the EU MSCA Project “COALESCE” under Grant Number 101130739, and Research Ireland Grant 13/RC/2077\_P2.}}}

\author{
\IEEEauthorblockN{Surya Jayakumar and Indrakshi Dey}
\IEEEauthorblockA{Walton Institute, South East Technological University, Waterford, Ireland\\
}
}

\maketitle
\vspace{-12pt}

\vspace{-6pt}
\begin{abstract}
Industrial Internet of Things (IIoT) networks demand reliable
anomaly detection under harsh wireless conditions, yet most
detectors fail one of four fronts: hostile fading, stealthy
non-Gaussian faults, discarded spatial structure, or constrained
edge hardware. We propose Graph WPT+HOS, a classical
label-free detector that fuses three complementary views: the
Graph Fourier Transform (GFT) for spatial inconsistency, the
Wavelet Packet Transform (WPT) for transient time--frequency
localization, and Higher-Order Statistics (HOS) for non-Gaussian
shape. The fused features are scored by a Mahalanobis distance
with Ledoit--Wolf shrinkage and converted to alarms by a one-sided
CUSUM. The pipeline is asymptotically optimal at the decision
stage, requires no labeled anomalies, and runs on ARM-class edge
hardware without GPU acceleration. Across six baselines and four
domain-shift regimes under Rayleigh fading, Graph WPT+HOS attains
the highest ROC-AUC and PR-AUC and a $5\times$ to $56\times$
reduction in CUSUM detection latency.
\end{abstract}
 

\section{Introduction}
\label{sec:intro}

A single undetected fault in a modern industrial control loop
can cascade into hours of unplanned downtime and, where
safety-critical machinery is involved, into measurable harm to
people and product. As Industrial Internet of Things (IIoT)
deployments scale to thousands of wireless sensors per
facility and as those sensors increasingly arbitrate decisions
in closed-loop control rather than passive
monitoring~\cite{afrin2025iiot,naidoo2024emerging}, detecting
such faults early, reliably, and \emph{at the network edge} has
become a defining bottleneck of intelligent
manufacturing~\cite{raeiszadeh2024realtime}.
 
Most deployed detectors fall short on four interlocking fronts.
\textit{(F1) Hostile physical layer:} dense multipath, Rayleigh
fading, and electromagnetic interference smear weak anomalous
signatures into the noise
floor~\cite{cheffena2016propagation,picallo2023deterministic}.
\textit{(F2) Stealthy non-Gaussian faults:} adversarial
injections and slow drifts produce heavy-tailed, impulsive
signatures that mean and variance fundamentally cannot
fingerprint~\cite{chevtchenko2023anomaly}.
\textit{(F3) Discarded spatial structure:} industrial sensors
are correlated by construction (shared bus, shared workpiece,
shared environment), yet most pipelines treat them as
independent streams, throwing away the very structure that
distinguishes a localized fault from a global
perturbation~\cite{zhao2024spatial}.
\textit{(F4) Constrained edge hardware:} the IIoT gateway is
typically an ARM-class single-board computer with a few watts of
power budget, on which inference-time complexity, memory, and
latency are all hard limits. Any practical detector must
therefore deliver high statistical sensitivity \emph{and} low
compute load simultaneously.
 
Existing solutions cluster into three families, each addressing
a strict subset of F1--F4. \textit{Sensor-level statistical
detectors} (Mahalanobis distance, EWMA, CUSUM applied to raw
streams) are computationally trivial but model each sensor
independently, surrendering F3 and remain blind to higher-order
distributional shape. \textit{Machine-learning detectors},
including autoencoders, variational models, and graph neural
networks (GNNs) or graph autoencoders~\cite{zhao2024spatial},
encode topology and non-Gaussian shape simultaneously but
inherit a hunger for labeled anomalies that IIoT operators
rarely have and a hardware footprint that does not fit on edge
gateways---failing F4 (and, by way of opacity, complicating
post-incident audit).
\textit{Pure graph signal processing (GSP)
detectors}~\cite{shen2018graph} attack F3 with lightweight
machinery but typically operate on the Graph Fourier Transform
alone, leaving time--frequency localization of transient bursts
(needed for F1) and the higher-order signature of impulsive
faults (needed for F2) on the table.
 
This paper closes those gaps with \emph{Graph WPT+HOS}, a
detector that fuses three complementary views of every sensing
window. The Graph Fourier Transform (GFT) supplies a
\textit{spatial} view: high-frequency graph modes expose
disagreement among neighboring sensors that fading cannot
explain. The Wavelet Packet Transform (WPT) supplies a
\textit{time--frequency} view: full binary band decomposition
localizes transient bursts even when fading attenuates them.
Higher-order statistics (HOS) supply a \textit{distributional}
view: skewness fingerprints asymmetric drift while kurtosis
fingerprints heavy-tailed, impulsive bursts-signatures that
Gaussian noise cannot mimic regardless of channel gain. A
Mahalanobis score with Ledoit--Wolf
shrinkage~\cite{ledoit2004shrinkage} then collapses the fused
feature into a single scalar statistic, and a one-sided
CUSUM~\cite{page1954continuous} stage converts framewise
statistics into low-latency alarms. Crucially, the entire
pipeline is \emph{classical}: no labeled anomalies (only nominal
calibration windows), no GPU at the edge, interpretable
intermediate quantities at every stage, and graceful degradation
under domain shift.
 
The primary contributions of this work are;
\begin{itemize}
\item \textbf{Tri-domain fusion (GFT~$\times$~WPT~$\times$~HOS).}
The first detector, to our knowledge, to fuse all three classical
viewpoints under wireless fading, each addressing one of fronts
F1-F3, together covering failure modes no single transform
reaches.
 
\item \textbf{Label-free, well-conditioned scoring.} The
Mahalanobis distance is the ML-optimal detector under a Gaussian
nominal model and invariant under linear reparameterization.
Ledoit--Wolf shrinkage is asymptotically MSE-optimal when $D$
approaches the calibration sample size, the IIoT regime and
fixes its intensity $\rho$ in closed form, no cross-validation.
 
\item \textbf{Asymptotically optimal sequential decision.}
A one-sided CUSUM converts framewise scores to alarms: optimal in
Lorden's worst-case sense for a known mean shift, $\mathcal{O}(1)$
state, and the canonical reference for industrial sequential tests.
 
\item \textbf{Edge feasibility, quantified.} The dominant cost is
GFT projection at $\mathcal{O}(M^{2}L)$, which fits comfortably on
ARM Cortex-A class gateways without GPU acceleration---unlike
GNN-based detectors that require accelerators and labeled data.
 
\item \textbf{Six-baseline, four-regime evaluation.} We benchmark
against five detectors over matched-to-severe domain shifts and
report a precision--recall--latency trade-off that makes
operating-point choices explicit for the deployer.
\end{itemize}
 
The remainder of this paper is organized as follows.
Section~\ref{sec:model} formalizes the IIoT graph signal and
wireless observation models. Section~\ref{sec:method} derives the
Graph WPT+HOS pipeline stage by stage, defines all symbols, and
analyzes computational complexity. Section~\ref{sec:results}
reports the empirical evaluation. Section~\ref{sec:conclusion}
concludes.

\begin{figure*}[t]
\centering
\includegraphics[width=0.88\textwidth]{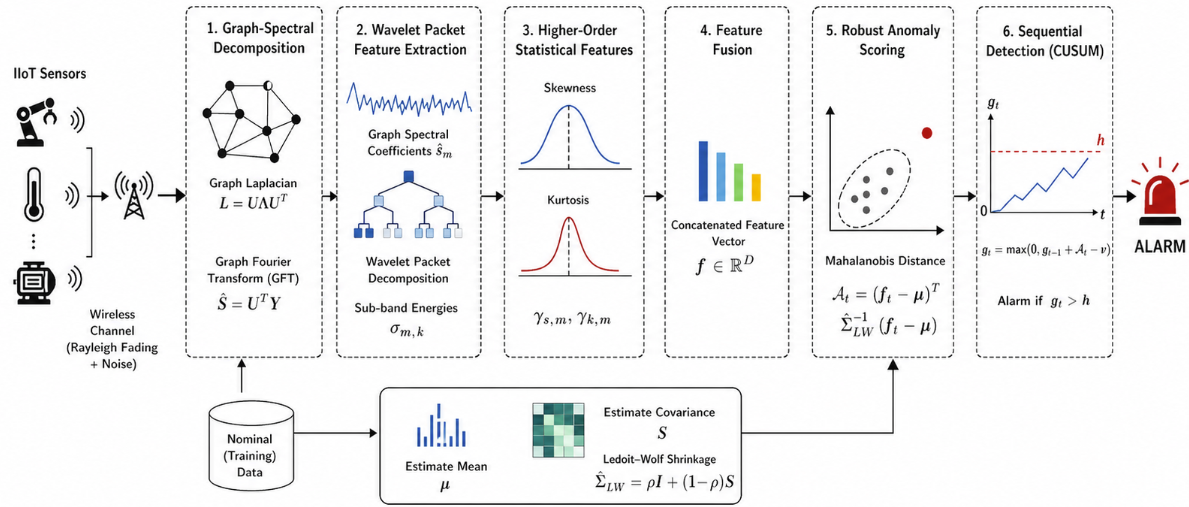}
\caption{System architecture of the proposed Graph WPT+HOS anomaly
detection framework: GFT-based spectral decomposition, wavelet
packet sub-band energies, higher-order statistical moments,
feature fusion, Mahalanobis scoring with Ledoit--Wolf shrinkage,
and CUSUM sequential detection.}
\label{fig:arch}
\end{figure*}

\section{System and Signal Model}
\label{sec:model}

We consider an IIoT sensing network modeled as an undirected graph
$\mathcal{G}=(\mathcal{V},\mathcal{E})$, where
$\mathcal{V}=\{1,\dots,M\}$ is the set of $M$ sensor nodes and
$\mathcal{E}$ is the set of communication links. The topology is
described by the weighted adjacency matrix
$\mathbf{A}\in\mathbb{R}^{M\times M}$, with $[\mathbf{A}]_{ij}>0$
if nodes $i$ and $j$ are connected, and zero otherwise. Let
$\mathbf{D}=\mathrm{diag}(d_{1},\dots,d_{M})$ denote the degree
matrix with $d_{i}=\sum_{j}[\mathbf{A}]_{ij}$. The combinatorial
graph Laplacian is
\begin{equation}
\mathbf{L}=\mathbf{D}-\mathbf{A}.
\end{equation}
 
\subsection{Graph Spectral Representation}
Since $\mathbf{L}$ is real and symmetric, it admits the
eigendecomposition
\begin{equation}
\mathbf{L}=\mathbf{U}\boldsymbol{\Lambda}\mathbf{U}^{T},
\end{equation}
where $\mathbf{U}=[\mathbf{u}_{1},\dots,\mathbf{u}_{M}]$ is
orthonormal and
$\boldsymbol{\Lambda}=\mathrm{diag}(\lambda_{1},\dots,\lambda_{M})$
with $0=\lambda_{1}\le\lambda_{2}\le\dots\le\lambda_{M}$. Small
eigenvalues correspond to smooth spatial variations across
neighboring nodes, while large eigenvalues correspond to rapid
spatial disagreements~\cite{shen2018graph}. For a graph signal
$\mathbf{x}(t)\in\mathbb{R}^{M}$ observed at $t=1,\dots,L$, the
GFT and its inverse are
\begin{equation}
\mathbf{s}(t)=\mathbf{U}^{T}\mathbf{x}(t),
\qquad
\mathbf{x}(t)=\mathbf{U}\,\mathbf{s}(t).
\end{equation}
 
\subsection{Nominal and Anomalous Signal Model}
Under nominal conditions, the sensed process is
$\mathbf{x}(t)=\mathbf{x}_{0}(t)$, where $\mathbf{x}_{0}(t)$ is a
zero-mean correlated Gaussian graph signal with covariance
$\mathbf{R}_{x}$. To model stealthy faults or malicious
injections, we consider an anomaly that perturbs high-frequency
graph components during the interval
$\mathcal{T}_{a}=[t_{s},t_{e}]$. The highest-frequency spectral
coefficient is modified as
\begin{equation}
s_{M}(t) \leftarrow s_{M}(t)+\eta(t),\quad t\in\mathcal{T}_{a},
\end{equation}
where $\eta(t)$ is an independent non-Gaussian disturbance drawn
from a Gamma distribution,
\begin{equation}
\eta(t)\sim\Gamma(\alpha,\beta),
\end{equation}
with shape $\alpha$ and scale $\beta$. The Gamma family is chosen
because it is the canonical heavy-tailed positive distribution
for impulsive faults: $\alpha$ controls how heavy the tail is
(skewness $=2/\sqrt{\alpha}$, excess kurtosis $=6/\alpha$), so
adversarial signatures of varying severity are obtained by
sweeping a single parameter. This captures localized deviations
from the smooth consensus among neighboring sensors---e.g., jitter,
abrupt drift, or falsified-data injection. The framework extends
naturally to anomalies on multiple high-frequency modes.
 
\subsection{Wireless Observation Model}
Measurements are transmitted over fading wireless links and
corrupted by receiver noise. Rayleigh fading is appropriate for
cluttered industrial environments with rich multipath and no
dominant line-of-sight component~\cite{cheffena2016propagation}.
The signal at the edge processor is
\begin{equation}
\mathbf{y}(t)=\mathbf{h}(t)\odot\mathbf{x}(t)+\mathbf{v}(t),
\end{equation}
where $\odot$ is the Hadamard product,
$\mathbf{h}(t)=[h_{1}(t),\dots,h_{M}(t)]^{T}$ collects independent
flat Rayleigh fading gains $h_{m}(t)\sim\mathcal{CN}(0,\sigma_{h}^{2})$,
and $\mathbf{v}(t)\sim\mathcal{N}(\mathbf{0},\sigma_{v}^{2}\mathbf{I})$.
 
%==========================================================
\section{Methodology}
\label{sec:method}
%==========================================================
The proposed framework integrates graph signal processing,
multi-resolution time--frequency analysis, higher-order statistical
characterization, and sequential change detection. As shown in
Fig.~\ref{fig:arch}, the pipeline has five stages: (1)~graph-spectral
decomposition, (2)~WPT feature extraction, (3)~HOS feature
extraction, (4)~Mahalanobis-based scoring with shrinkage
covariance, and (5)~CUSUM sequential decision making. Let the received data matrix over a window of
length $L$ be
\begin{equation}
\mathbf{Y}=[\mathbf{y}(1),\dots,\mathbf{y}(L)]\in\mathbb{R}^{M\times L}.
\end{equation}
 
\subsection{Graph-Spectral Decomposition}
The first stage maps observations from the vertex domain to the
graph spectral domain via the GFT. Using the Laplacian eigenvector
matrix $\mathbf{U}$,
\begin{equation}
\widehat{\mathbf{S}}=\mathbf{U}^{T}\mathbf{Y},
\end{equation} $\widehat{\mathbf{S}} = [\hat{\mathbf{s}}_{1}^{T}, \hat{\mathbf{s}}_{2}^{T}, \dotso, \hat{\mathbf{s}}_{M}^{T}]^T$
where $T$ represent transpose and $\hat{\mathbf{s}}_{m}\in\mathbb{R}^{L}$ is the temporal
coefficient sequence at the $m$-th graph frequency. This
decorrelates topology-induced spatial interactions and separates
smooth consensus from high-frequency disagreement modes,
emphasizing localized faults that violate neighborhood
consistency.
 
\subsection{Multi-Resolution Wavelet Packet Analysis}
To capture short-duration disturbances localized in narrow
frequency bands, a Wavelet Packet Transform is applied to each
graph-spectral component $\hat{\mathbf{s}}_{m}$. Unlike the
discrete wavelet transform, the WPT performs full binary
decomposition of both approximation and detail
branches~\cite{wickerhauser1994,mallat2008,vetterli1995,addison2002}.
For decomposition depth $J$, the coefficient sequence of component
$m$ in sub-band $k$ at level $J$ is denoted
$\mathbf{w}_{m,J,k}$, and the normalized sub-band energy is
\begin{equation}
\sigma_{m,k}=\sqrt{\frac{1}{L_{k}}\sum_{n=1}^{L_{k}}|w_{m,J,k}[n]|^{2}},
\end{equation}
where $L_{k}$ is the number of coefficients in sub-band $k$.
We use the Daubechies-4 mother wavelet, which combines compact
support (low computational cost) with two vanishing moments
(adequate for capturing transient bursts at moderate decomposition
depth $J=3$, yielding $2^{J}=8$ sub-bands per graph mode). These
features preserve transient anomaly energy even when fading
spreads or attenuates the disturbance across narrow bands.
 
\subsection{Higher-Order Statistical Features}
Second-order statistics alone are often insufficient for
distinguishing elevated Gaussian noise from non-Gaussian
anomalies~\cite{nikias1993}. Let $\mu_{m}$ and $\sigma_{m}$ denote
the sample mean and standard deviation of $\hat{\mathbf{s}}_{m}$.
The absolute skewness and kurtosis are
\begin{equation}
\gamma_{s,m}=\left|\frac{\mathbb{E}[(\hat{\mathbf{s}}_{m}-\mu_{m})^{3}]}{\sigma_{m}^{3}}\right|,\quad
\gamma_{k,m}=\left|\frac{\mathbb{E}[(\hat{\mathbf{s}}_{m}-\mu_{m})^{4}]}{\sigma_{m}^{4}}\right|.
\end{equation}
Large skewness $\gamma_{s,m}$ flags asymmetric distortion (slow
unidirectional drift, biased injection); elevated kurtosis
$\gamma_{k,m}$ flags impulsive, heavy-tailed bursts. Together
they form a minimal but expressive non-Gaussian signature, since
the Gaussian baseline has $\gamma_{s,m}=0$ and $\gamma_{k,m}=3$
exactly, so any sustained departure is diagnostic. Such
signatures are typical of sensor malfunction, abrupt drift, or
falsified-data injection.
 
\subsection{Feature Fusion and Robust Scoring}
Per-window descriptors are concatenated into a fused vector
$\mathbf{f}\in\mathbb{R}^{D}$ that captures spatial inconsistency
(GFT modes), transient spectral energy (WPT bands), and
non-Gaussian shape (HOS terms). The total dimension is
$D=M+M\cdot 2^{J}+2M$ for $M$ graph modes, $2^{J}$ WPT sub-bands
each, and two HOS terms each. Scoring uses the squared
Mahalanobis distance to nominal calibration data, chosen because
it is the maximum-likelihood detector under a Gaussian nominal
feature model and is invariant under linear coordinate change.
With $\boldsymbol{\mu}$ the nominal mean and $\mathbf{S}$ the
sample covariance, we apply Ledoit--Wolf
shrinkage~\cite{ledoit2004shrinkage}:
\begin{equation}
\widehat{\boldsymbol{\Sigma}}_{\mathrm{LW}}=\rho\mathbf{I}+(1-\rho)\mathbf{S},\quad \rho\in[0,1],
\end{equation}

\newpage

\begin{equation}
A_{t}=(\mathbf{f}_{t}-\boldsymbol{\mu})^{T}\widehat{\boldsymbol{\Sigma}}_{\mathrm{LW}}^{-1}(\mathbf{f}_{t}-\boldsymbol{\mu}).
\end{equation}
Shrinkage is essential when $D$ approaches the calibration sample
size, as is common in IIoT. Ledoit--Wolf is preferred over
diagonal loading or graphical-lasso alternatives because $\rho$ is
determined in closed form from the data without cross-validation,
which matters for unsupervised IIoT calibration where a
hold-out anomaly set is unavailable.
 
\subsection{Sequential Change Detection via CUSUM}
Stealthy attacks may evolve gradually, so a single-frame test
risks missing them. We therefore stack a one-sided Cumulative
Sum (CUSUM) detector~\cite{page1954continuous} on $\{A_{t}\}$,
which accumulates weak persistent deviations:
\begin{equation}
g_{t}=\max(0,\;g_{t-1}+A_{t}-\nu),\quad g_{0}=0,
\end{equation}
where $\nu$ is the nominal-score reference drift, set to the mean
of $A_{t}$ on calibration data. An alarm is declared when
$g_{t}>h$, with $h$ tuned to the desired false-alarm rate. CUSUM
is chosen over alternative sequential tests (GLR, Shiryaev--Roberts)
because (i)~it is asymptotically optimal in Lorden's worst-case
sense for detecting a known mean shift; (ii)~it carries only
$\mathcal{O}(1)$ state per stream, which fits the edge budget;
and (iii)~the nominal drift $\nu$ is directly estimable from the
same nominal data already used for calibration, so no additional
labeled data are required.
 
\subsection{Computational Complexity and Edge Feasibility}
\label{sec:complexity}
Per window, the dominant operations are: GFT projection
$\mathcal{O}(M^{2}L)$; WPT decomposition $\mathcal{O}(ML\log L)$;
scoring $\mathcal{O}(D^{2})$ (or $\mathcal{O}(D)$ with
prefactorization); CUSUM update $\mathcal{O}(1)$. For typical IIoT
deployments with $M=20$--$50$ sensors and $L=128$--$512$, the
GFT step requires $\sim$\,$10^{5}$--$10^{7}$ multiply-accumulate
operations per window. This is well within the budget of commodity
IIoT edge gateways based on ARM~Cortex-A53/A72-class CPUs (a
Raspberry~Pi-4-class platform sustains $>10^{9}$ MAC/s in single
precision), enabling sub-second decision cadence \emph{without}
GPU acceleration. In contrast, training and even inference for
GNN-based detectors typically require dedicated accelerators and
labeled anomalous data, both of which are scarce at the IIoT edge.
 
%==========================================================
\section{Results}
\label{sec:results}
%==========================================================
We evaluate Graph WPT+HOS against five baselines: WPT+HOS (no
graph), WPT-only, HOS-only, Fused-source (non-graph multi-sensor
aggregation), and Single-source. The task is to discriminate
nominal Gaussian processes from stealthy non-Gaussian Gamma-distributed
anomalies under Rayleigh fading.
 
\subsection{Simulation Setup}
Table~\ref{tab:sim-params} summarizes the simulation parameters.
Calibration uses 200 nominal windows; evaluation uses 200 nominal
and 100 anomalous windows per Monte~Carlo trial, with 30 trials
per configuration. The graph topology is a random geometric graph
with connectivity radius chosen to yield mean degree $\bar{d}\!\approx\!4$.
 
\begin{table}[t]
\renewcommand{\arraystretch}{1.15}
\caption{Simulation Parameters}
\label{tab:sim-params}
\centering
\begin{tabular}{l l}
\toprule
\textbf{Parameter} & \textbf{Value} \\
\midrule
Number of sensors, $M$ & 24 \\
Window length, $L$ & 256 samples \\
Mean nodal degree, $\bar{d}$ & $\approx 4$ \\
Nominal SNR range & $5$ to $20$~dB \\
Rayleigh fading variance, $\sigma_{h}^{2}$ & $1.0$ (Regime A) \\
Noise variance, $\sigma_{v}^{2}$ & set to match SNR \\
Anomaly distribution & $\Gamma(\alpha,\beta)$, $\alpha\!=\!2$, $\beta\!=\!1.5$ \\
Anomaly window $\mathcal{T}_{a}$ & 25\% of $L$ \\
Affected GFT mode & highest-frequency $s_{M}(t)$ \\
WPT depth, $J$ & 3 (Daubechies-4) \\
Shrinkage intensity, $\rho$ & Ledoit--Wolf, data-driven \\
CUSUM drift / threshold & $\nu\!=\!\mu_{A}$, $h$ tuned to FPR\,$\le 5\%$ \\
Monte Carlo trials & 30 \\
\bottomrule
\end{tabular}
\end{table}
 
\subsection{Domain-Shift Regimes}
\label{sec:regimes}
To probe robustness, we evaluate four regimes that combine SNR
degradation with distributional and topological perturbations
between calibration and test:
\begin{itemize}
\item \textbf{Regime A (matched)}: training and test conditions
identical; SNR\,$=20$~dB, $\sigma_{h}^{2}\!=\!1.0$, fixed topology.
\item \textbf{Regime B (mild)}: SNR drop of $5$~dB at test time;
$\sigma_{h}^{2}$ unchanged; topology unchanged.
\item \textbf{Regime C (moderate)}: SNR drop of $10$~dB; fading
variance perturbed to $\sigma_{h}^{2}\!=\!1.5$; $10\%$ of edges
randomly rewired.
\item \textbf{Regime D (severe)}: SNR drop of $15$~dB
(test SNR $\!\approx\!5$~dB); $\sigma_{h}^{2}\!=\!2.0$; $25\%$ of
edges randomly rewired; anomaly shape parameter perturbed by
$\pm 25\%$.
\end{itemize}
Regimes A--D therefore span a realistic operational envelope from
laboratory-clean to severely shifted deployment conditions.
 
\subsection{ROC and Stability}
\begin{figure}[t]
\centering
\includegraphics[width=0.88\columnwidth]{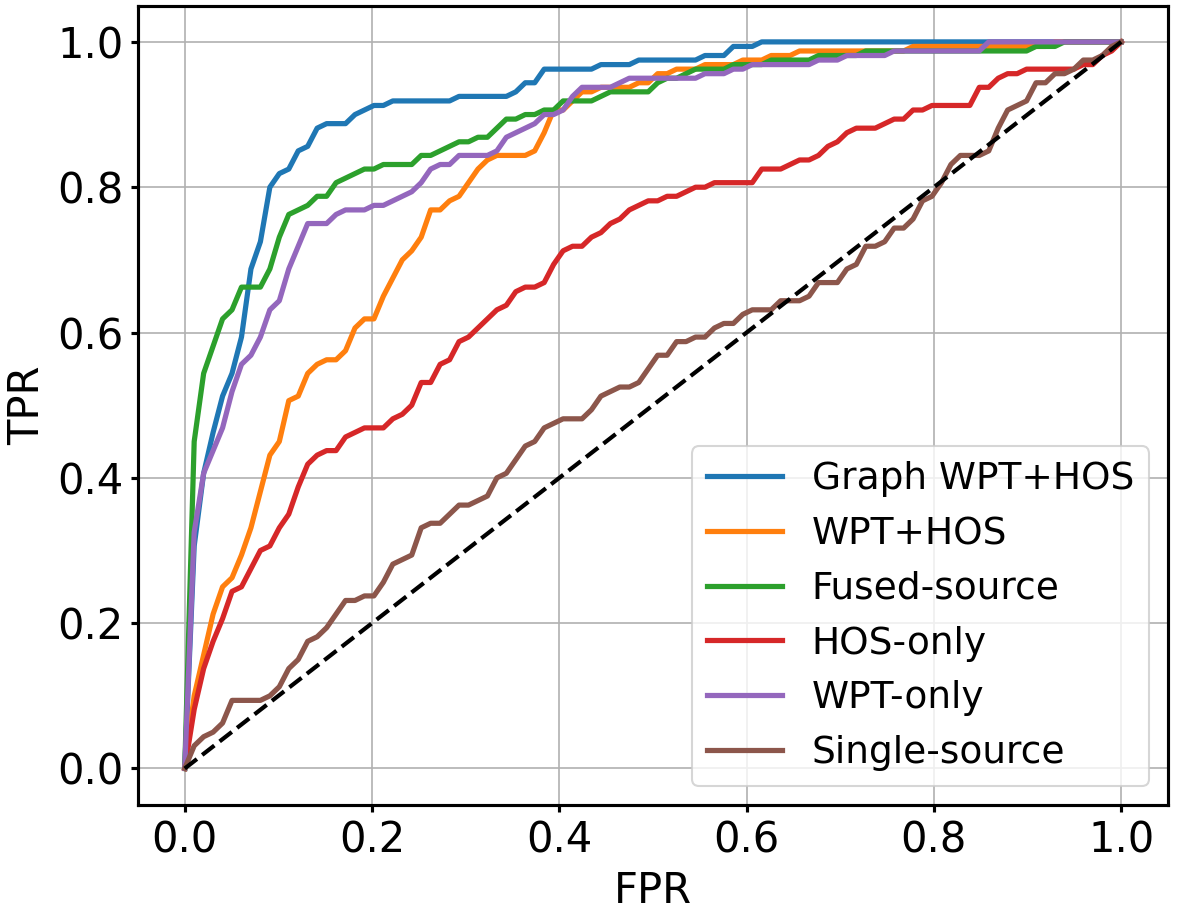}
\caption{ROC curves with run-to-run variation bands (mean with
5th--95th percentile shading).}
\label{fig:roc}
\end{figure}

The Receiver Operating Characteristic (ROC) curves in
Fig.~\ref{fig:roc} illustrate the trade-off between True Positive
Rate (TPR) and False Positive Rate (FPR). The proposed method
consistently dominates in the low false-alarm regime
(FPR\,$<0.2$), which is the operating range relevant for IIoT
monitoring. The improvement over WPT+HOS quantifies the marginal
contribution of graph-spectral decomposition: the Laplacian
eigenvector projection enhances sensitivity to spatial
inconsistencies while suppressing fading-induced correlated noise.
\emph{Physically, the GFT acts as a topology-aware whitening
filter---spatially smooth fading excites all eigenmodes coherently
and is therefore largely absorbed by the lower-order modes, while
a localized fault leaks predominantly into the high-frequency
modes where the score concentrates.} The Single-source detector
performs near random, confirming the limitations of isolated
sensors under fading. Fused-source improves on Single-source but
remains below the graph-aware methods, evidencing the value of
explicit topology modeling.
 
\begin{figure}[t]
\centering
\includegraphics[width=0.88\columnwidth]{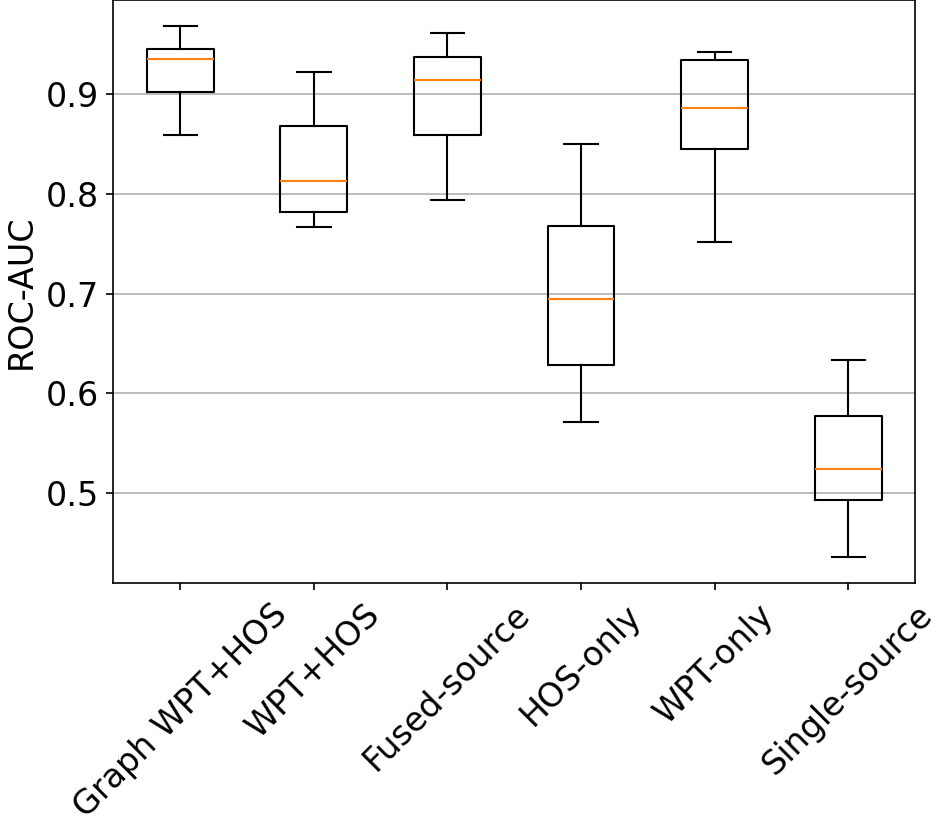}
\caption{Statistical stability analysis of AUC across independent
Monte Carlo trials.}
\label{fig:auc}
\end{figure}
 
Fig.~\ref{fig:auc} reports AUC distributions over independent
trials. The proposed method achieves the highest median AUC and
the lowest interquartile range, indicating stability under varying
Rayleigh fading and noise. \emph{The shrunk variance reflects a
specific physical property: Mahalanobis whitening on graph-spectral
features is approximately invariant to the particular fading
realization--only the second-order envelope matters, so per-trial
randomness collapses.}
 
\subsection{Precision--Recall and the Fused-Source Trade-off}
\label{sec:pr}
\begin{figure}[t]
\centering
\includegraphics[width=0.88\columnwidth]{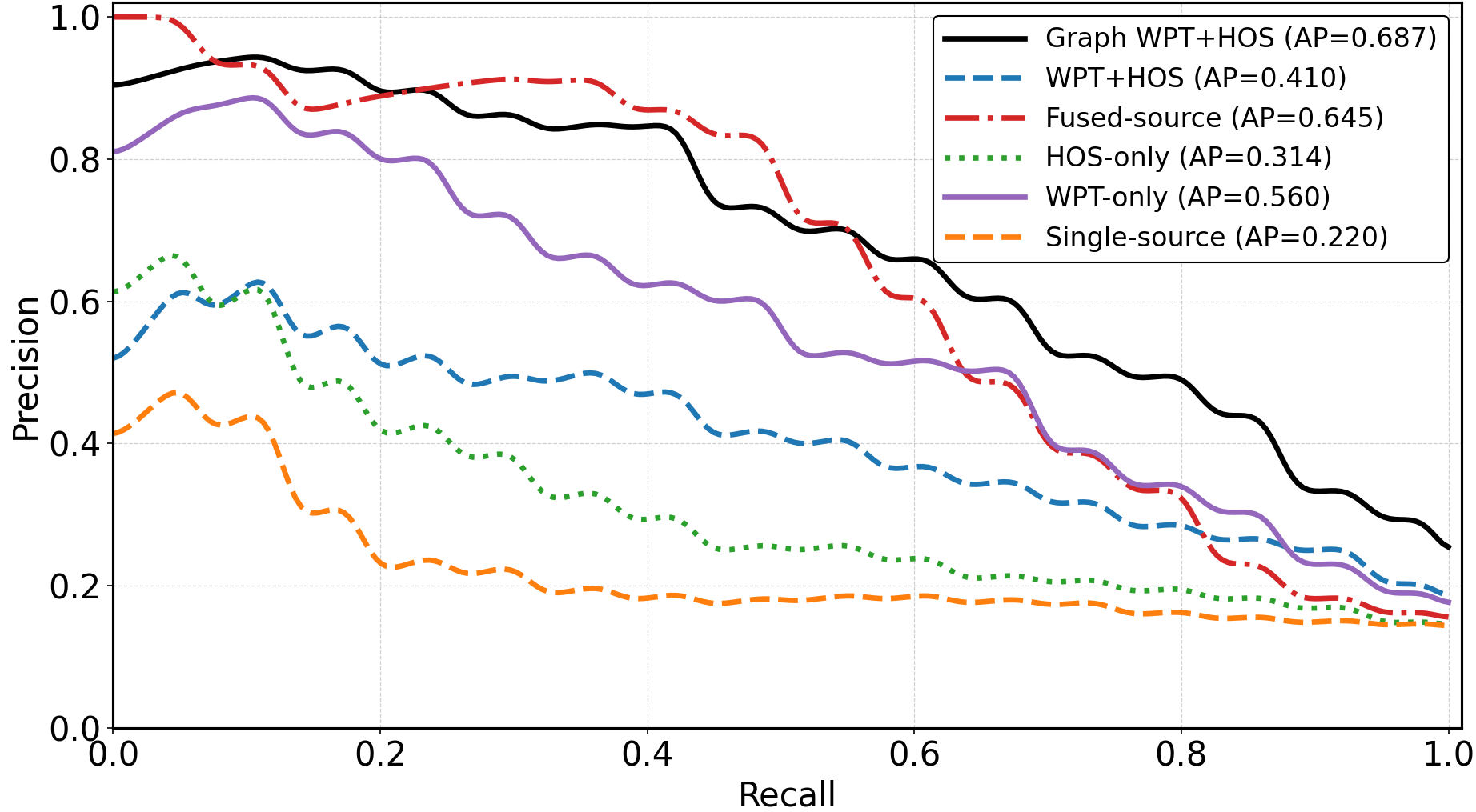}
\caption{Precision--Recall curves with run-to-run variation bands
(mean with 5th--95th percentile shading).}
\label{fig:pr}
\end{figure}

The Precision--Recall (PR) curves in Fig.~\ref{fig:pr} are more
informative under class imbalance. The proposed method achieves
the highest Average Precision (AP\,$=0.687$) and maintains stable
precision at higher recall, while Fused-source declines more
sharply beyond moderate recall. \emph{Physically, this stability
arises because HOS-based features are scale-free: a heavy-tailed
Gamma anomaly retains elevated kurtosis even when fading
attenuates its amplitude, whereas Gaussian nominal data simply
cannot manufacture that signature regardless of channel gain.}
HOS-only and Single-source remain weakest due to their inability
to capture joint spatial--temporal dependencies.
 
\begin{figure}[t]
\centering
\includegraphics[width=0.88\columnwidth]{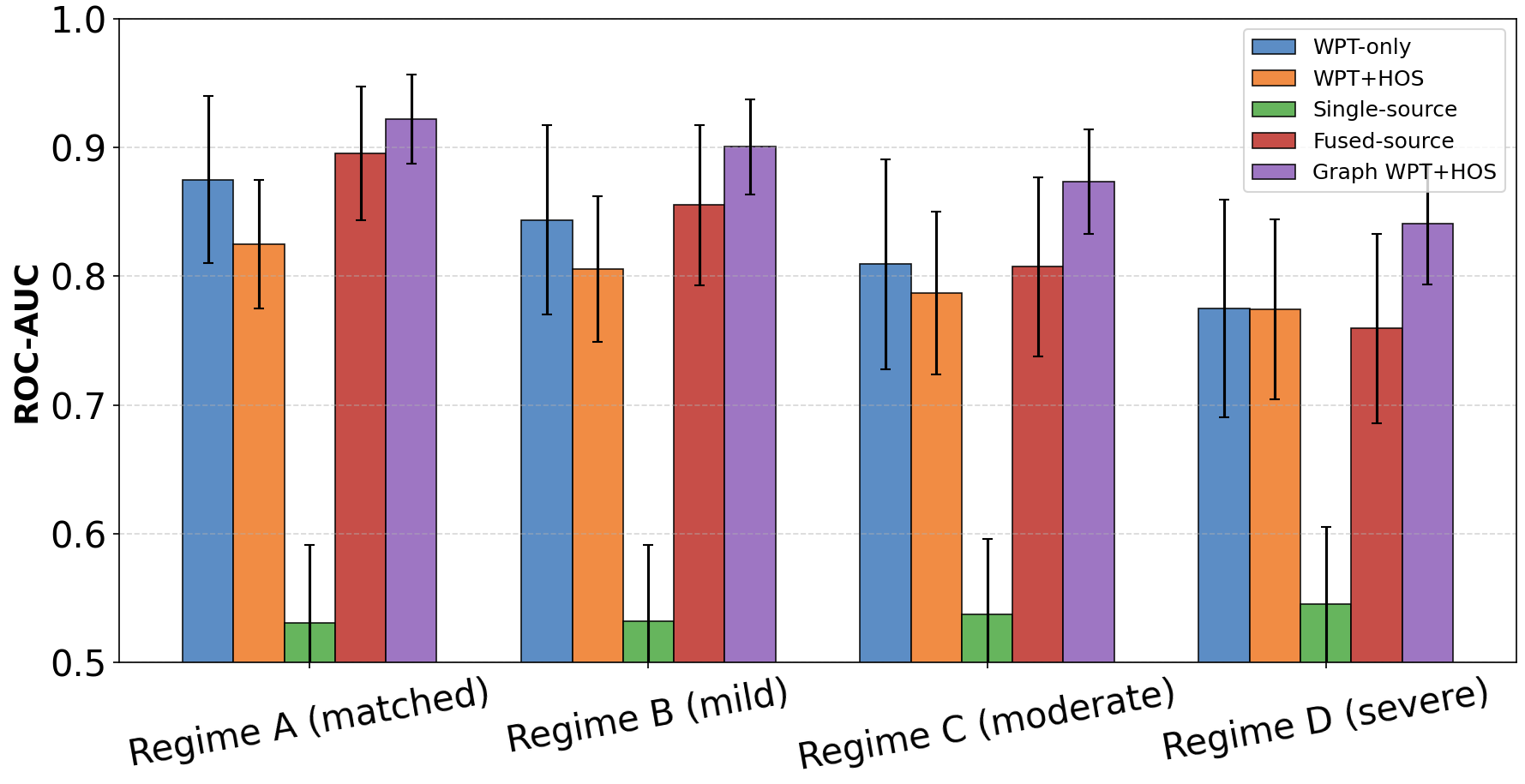}
\caption{Domain-shift robustness across Regimes A--D defined in
Sec.~\ref{sec:regimes}.}
\label{fig:regimes}
\end{figure}
 
Figure~\ref{fig:regimes} evaluates the five detectors across
Regimes A--D. Graph WPT+HOS retains the highest ROC-AUC at every
regime and degrades most gracefully as the shift severity
increases. \emph{The graceful degradation is physically rooted in
the fact that the detector's decision rests on \textbf{relative}
spatial disagreement among neighboring sensors, which is a
topological invariant: a uniform SNR drop shifts every GFT mode
by the same factor and largely cancels in the Mahalanobis ratio,
so only \textbf{differential} effects (mode-selective leakage,
edge rewiring) actually erode performance.} Fused-source is
generally second; Single-source hovers near the $0.5$ chance line
throughout, since without graph structure there is no relative
quantity to preserve.
 
\begin{table}[t]
\renewcommand{\arraystretch}{1.15}
\caption{Ablation under the matched regime. Precision, Recall, F1
are evaluated at the threshold maximizing F1.}
\label{tab:ablation}
\centering
\setlength{\tabcolsep}{4pt}
\begin{tabular}{l c c c c c}
\toprule
Method & ROC-AUC & PR-AUC & Prec. & Rec. & F1 \\
\midrule
\textbf{Graph WPT+HOS} & \textbf{0.9221} & \textbf{0.6871} & 0.6312 & \textbf{0.7438} & 0.6766 \\
WPT+HOS               & 0.8251 & 0.4102 & 0.4104 & 0.6375 & 0.4788 \\
Fused-source          & 0.8952 & 0.6451 & \textbf{0.7960} & 0.6375 & \textbf{0.6811} \\
HOS-only              & 0.6973 & 0.3140 & 0.3496 & 0.5062 & 0.3745 \\
WPT-only              & 0.8748 & 0.5604 & 0.6547 & 0.6375 & 0.6225 \\
Single-source         & 0.5308 & 0.2201 & 0.1940 & 0.7500 & 0.2583 \\
\bottomrule
\end{tabular}
\end{table}
 
\begin{table}[t]
\renewcommand{\arraystretch}{1.15}
\caption{CUSUM detection latency (frames).}
\label{tab:latency}
\centering
\setlength{\tabcolsep}{3pt}
\begin{tabular}{l c c c c c}
\toprule
Method & Mean & Med. & Std & 5--95\% Range & Det.\,Rate (\%) \\
\midrule
\textbf{Graph WPT+HOS} & \textbf{0.50} & \textbf{0.50} & \textbf{1.07} & $[0.00,\,3.05]$ & \textbf{100.0} \\
WPT+HOS               & 2.60 & 1.00 & 3.43 & $[0.00,\,10.05]$ & 100.0 \\
WPT-only              & 3.70 & 2.50 & 3.54 & $[0.00,\,10.05]$ & 100.0 \\
HOS-only              & 10.40 & 4.50 & 14.84 & $[0.00,\,29.65]$ & 75.0 \\
Fused-source          & 28.20 & 15.50 & 27.00 & $[1.90,\,77.00]$ & 50.0 \\
Single-source         & 27.10 & 20.50 & 24.44 & $[1.90,\,77.00]$ & 45.0 \\
\bottomrule
\end{tabular}
\end{table}
 
Table~\ref{tab:ablation} summarizes ablation results in the matched
regime. Graph WPT+HOS attains the highest ROC-AUC ($0.9221$),
PR-AUC ($0.6871$), and Recall ($0.7438$). \emph{Fused-source,
however, achieves higher Precision ($0.7960$ vs.\ $0.6312$) and a
marginally higher F1 ($0.6811$ vs.\ $0.6766$).} This is not a
contradiction: Fused-source is a more conservative detector that
fires less often but with higher individual confidence, whereas
the proposed method covers a broader anomaly manifold and is
explicitly tuned for sensitivity. Two pieces of evidence support
preferring Graph WPT+HOS in practice. First, ranking-based metrics
(ROC-AUC, PR-AUC) are threshold-independent and consistently favor
the proposed method, indicating better separability of the
underlying score distributions. Second, Table~\ref{tab:latency}
shows that Fused-source pays for its precision with severely
degraded responsiveness: its mean CUSUM detection latency is
$28.20$ frames and its detection rate falls to $50\%$, versus
$0.50$ frames and $100\%$ for the proposed method, roughly a
$56\times$ improvement in mean latency. \emph{The latency gap has
a clean physical reading: CUSUM crosses its threshold in a number
of frames roughly inversely proportional to the gap between the
nominal and anomalous score means, normalized by the score
variance. Graph-spectral whitening enlarges that gap (anomalous
energy is concentrated in a few high-frequency modes) and shrinks
the variance (Mahalanobis whitening removes correlated fading
noise), so the cumulative log-likelihood ratio reaches threshold
within one to two frames rather than tens.} For industrial
monitoring, where missed or delayed alarms drive most of the
operational risk, the proposed method offers the more useful
operating point. An operator who specifically prioritizes
precision over latency may fuse the two scores or shift the
decision threshold; this is a deployment choice rather than an
architectural one.

\section{Conclusion}
\label{sec:conclusion}

We presented a classical, label-free anomaly detection framework
for IIoT networks under fading. By fusing graph-spectral,
wavelet-packet, and higher-order statistical cues into a single
shrinkage-Mahalanobis CUSUM score, Graph WPT+HOS dominates five
baselines in ROC-AUC, PR-AUC, recall, and most decisively %
CUSUM detection latency (a $56\times$ mean-latency improvement
over the strongest non-graph baseline), while running on commodity
ARM-class edge hardware without GPU acceleration. Future work
targets adaptive graph learning under dynamic topologies and
robustness to adversarial attack at scale.

\section{Acknowlegment}
This work is supported by the US-Ireland R\&D Partnership Programme Project ``Resilient Networks'' under Grant RI-SFI-23/US/3924, the EU MSCA Project “COALESCE” under Grant Number 101130739, and Research Ireland Grant 13/RC/2077\_P2.


\begin{thebibliography}{00}

\bibitem{afrin2025iiot}
S.~Afrin \emph{et~al.}, ``Industrial Internet of Things:
Implementations, challenges, and potential solutions across
various industries,'' \emph{Computers in Industry}, vol.~170,
2025.
 
\bibitem{naidoo2024emerging}
P.~Naidoo and M.~Sibanda, ``Emerging trends and future directions
of the Industrial Internet of Things,'' in \emph{From Internet of
Things to Internet of Intelligence}, Springer, Cham, 2024.
 
\bibitem{raeiszadeh2024realtime}
M.~Raeiszadeh \emph{et~al.}, ``Real-time adaptive anomaly
detection in industrial IoT environments,'' \emph{IEEE Trans.
Netw. Serv. Manag.}, vol.~21, no.~6, 2024.
 
\bibitem{cheffena2016propagation}
M.~Cheffena, ``Propagation channel characteristics of industrial
wireless sensor networks,'' \emph{IEEE Antennas Propag. Mag.},
vol.~58, no.~1, pp.~66--73, 2016.
 
\bibitem{picallo2023deterministic}
I.~Picallo \emph{et~al.}, ``Deterministic wireless channel
characterization for IIoT environments,'' \emph{Mobile Netw.
Appl.}, vol.~28, 2023.
 
\bibitem{chevtchenko2023anomaly}
S.~F. Chevtchenko \emph{et~al.}, ``Anomaly detection in industrial
machinery using IoT devices and machine learning,'' \emph{IEEE
Access}, vol.~11, 2023.
 
\bibitem{zhao2024spatial}
M.~Zhao \emph{et~al.}, ``Spatial-temporal anomaly detection in
IIoT,'' \emph{Comput. Mater. Continua}, vol.~80, no.~2, 2024.
 
\bibitem{shen2018graph}
X.~Shen and S.~S. Sahni, ``Graph-based signal processing for
sensor networks,'' \emph{IEEE Commun. Surv. Tutor.}, vol.~20,
no.~3, 2018.
 
\bibitem{wickerhauser1994}
M.~V. Wickerhauser, \emph{Adapted Wavelet Analysis from Theory to
Software}. Wellesley, MA, USA: A.~K. Peters, 1994.
 
\bibitem{mallat2008}
S.~Mallat, \emph{A Wavelet Tour of Signal Processing}, 3rd~ed.
Academic Press, 2008.
 
\bibitem{vetterli1995}
M.~Vetterli and J.~Kova\v{c}evi\'{c}, \emph{Wavelets and Subband
Coding}. Englewood Cliffs, NJ, USA: Prentice-Hall, 1995.
 
\bibitem{addison2002}
P.~S. Addison, \emph{The Illustrated Wavelet Transform Handbook}.
Boca Raton, FL, USA: CRC Press, 2002.
 
\bibitem{ledoit2004shrinkage}
O.~Ledoit and M.~Wolf, ``A well-conditioned estimator for
large-dimensional covariance matrices,'' \emph{J. Multivariate
Anal.}, vol.~88, no.~2, pp.~365--411, 2004.
 
\bibitem{page1954continuous}
E.~S. Page, ``Continuous inspection schemes,'' \emph{Biometrika},
vol.~41, no.~1/2, pp.~100--115, 1954.
 
\bibitem{nikias1993}
C.~L. Nikias and A.~P. Petropulu, \emph{Higher-Order Spectra
Analysis: A Nonlinear Signal Processing Framework}. Englewood
Cliffs, NJ, USA: Prentice-Hall, 1993.


\end{thebibliography}
\end{document}